\documentclass[graybox]{svmult}

\usepackage{mathptmx}       
\usepackage{helvet}         
\usepackage{courier}        
\usepackage{type1cm}        
\usepackage{makeidx}         
\usepackage{graphicx}        
\usepackage{multicol}        
\usepackage[bottom]{footmisc}

\makeindex             

\newcommand{\be}{\begin{equation}}
\newcommand{\ee}{\end{equation}}
\newcommand{\bra}{\langle}
\newcommand{\ket}{\rangle}
\newcommand{\bea}{\begin{eqnarray}}
\newcommand{\eea}{\end{eqnarray}}
\newcommand{\dis}{\displaystyle}

\begin{document}

\title*{Bayesian estimation of GARCH model with an adaptive proposal density}

\author{Tetsuya Takaishi}%

\institute{Tetsuya Takaishi \at Hiroshima University of Economics,
\email{takaishi@hiroshima-u.ac.jp}
}

\maketitle

\abstract{
A Bayesian estimation of a GARCH model is performed 
for US Dollar/Japanese Yen exchange rate by the Metropolis-Hastings algorithm 
with a proposal density given by the adaptive construction scheme.
In the adaptive construction scheme the proposal density is
assumed to take a form of a multivariate Student's t-distribution 
and its parameters are evaluated by using the sampled data 
and updated adaptively during Markov Chain
Monte Carlo simulations. 
We find that the autocorrelation times between the data 
sampled by the adaptive construction scheme
are considerably reduced. 
We conclude that the adaptive construction scheme works efficiently for the Bayesian inference of the GARCH model. 
}


\section{Introduction}
In finance volatility of asset returns plays an important role to manage financial risk.
To forecast volatility, 
various empirical models which mimic the properties of the volatility
have been proposed.
Engle\cite{ARCH} proposed Autoregressive Conditional Heteroskedasticity (ARCH) model
where the present volatility is assumed to depend on the squares of past observations.
Later Bollerslev\cite{GARCH} advocated Generalized ARCH (GARCH) model which 
is an extension of the ARCH model and includes additional past volatility terms to the present volatility estimate. 
It is known that the volatility of the financial assets
exhibits clustering in the financial time series.
The GARCH model can captures this property. 
Furthermore the return distribution generated from the GARCH process shows a fat-tailed distribution 
which is also seen in the real financial markets. 
There also exists extension of the GARCH model which incorporates 
the asymmetric property of the volatility\cite{EGARCH,GJR,APGARCH,QGARCH}.

A preferred algorithm to infer GARCH model parameters is the Maximum Likelihood (ML) method
which estimates the parameters by maximaizing the corresponding likelihood function of the GARCH model.
In this algorithm there is a practical difficulty in the maximization procedure when the output results are sensitive to  
starting values.

By the recent computer development the Bayesian inference by Markov chain Monte Carlo (MCMC) methods, 
which is an alternative approach to estimate GARCH parameters, 
has become popular.
There exist a variety of methods proposed to implement the MCMC scheme\cite{Bauwens}-\cite{HMC}. 
In a recent survey\cite{ASAI} it is shown that Acceptance-Rejection/Metropolis-Hastings  (AR/MH) algorithm 
works better than other algorithms.
In the AR/MH algorithm the proposal density is assumed to be a multivariate Student's t-distribution and 
the parameters to specify the distribution are estimated by the ML technique.
Recently a new method to estimate those parameters without relying on the ML technique was proposed\cite{ACS}.
In the method  the parameters are determined by an MCMC simulation.
During the MCMC simulation, the parameters are updated adaptively using the data sampled by the MCMC method itself.
We call this method "adaptive construction scheme".
The adaptive construction scheme was tested for artificial GARCH data and 
it is shown that the adaptive construction scheme can significantly reduce 
the correlation between sampled data\cite{ACS}.  
In this study we apply the adaptive construction scheme to 
real financial data, US Dollar/Japanese Yen exchange rate 
and examine the efficiency of the adaptive construction scheme. 

\section{GARCH Model}

The GARCH(p,q) model by Bollerslev\cite{GARCH} is given by
\be
y_t=\sigma_t \epsilon_t ,
\ee
\be
\sigma_t^2  = \omega + \sum_{i=1}^{q}\alpha_i y_{t-i}^2 
+ \sum_{i=1}^{p}\beta_i \sigma_{t-i}^2,
\ee
where the GARCH parameters are restricted to $\omega>0$, $\alpha_i>0$ and $\beta_i>0$ to ensure a positive volatility,
and the stationary condition $\sum_{i=1}^{q}\alpha_i + \sum_{i=1}^{p}\beta_i <1$ is also required.
$\epsilon_t$ is an independent normal error $\sim N(0,1)$.

In this study  we focus on GARCH(1,1) model 
where the  volatility $\sigma_t^2$ is given by  
\be 
\sigma_t^2  = \omega + \alpha y_{t-1}^2 + \beta \sigma_{t-1}^2.
\ee
The likelihood function of the GARCH model is given by
\be
L(y|\theta)=\Pi_{i=1}^{n} \frac1{\sqrt{2\pi\sigma_t^2}}\exp\left.(-\frac{y_t^2}{\sigma_t^2}\right.).
\ee

\section{Bayesian inference}
Using  Bayes' rule  
the posterior density $\pi(\theta|y)$  
with $n$ observations denoted by $y=(y_1,y_2,\dots,y_n)$ is given by
\be
\pi(\theta|y)\propto L(y|\theta) \pi(\theta),
\ee
where $L(y|\theta)$ is the likelihood function.
$\pi(\theta)$ is the prior density which we have to specify 
depending on $\theta$.
In this study we assume that the prior density $\pi(\theta)$ is constant.

With $\pi(\theta|y)$ we infer $\theta$ as expectation values of
$\theta$. 
The expectation values are given by
\be
\bra {\bf \theta} \ket = \frac1{Z}\int {\bf \theta} \pi(\theta|y) d\theta,
\label{eq:int}
\ee
where $Z=\int \pi(\theta|y) d\theta$ is the normalization constant. 
Hereafter we omit $Z$ since this factor is irrelevant to MCMC estimations.

The MCMC technique gives a method to estimate eq.(\ref{eq:int}) numerically.
The basic procedure of the MCMC method is as follows.
First we sample $\theta$ drawn from a probability distribution
$\pi(\theta|y)$. 
Sampling is done by a technique which produces a Markov chain.  
After sampling  some data, 
we evaluate the expectation value as an average value over the sampled data $\theta^{(i)}$, 
\be
\bra {\bf \theta} \ket = \lim_{k \rightarrow \infty} \frac1k\sum_{i=1}^k \theta^{(i)},
\ee
where 
$k$ is the number of the sampled data.
The statistical error for $k$ independent data 
is proportional to $\frac1{\sqrt{k}}$.
When the sampled data are correlated 
the statistical error will be proportional to $\sqrt{\frac{2\tau}{k}}$ 
where $\tau$ is the autocorrelation time between the sampled data.
The autocorrelation time depends on the MCMC method we employ.
Thus it is desirable to take an MCMC method which 
can generate data with a small $\tau$.

\section{Metropolis-Hastings algorithm}

The Metropolis-Hastings (MH) algorithm\cite{MH} is  an MCMC simulation method which generates
draws from any probability density. 
The MH algorithm is an extension of the original Metropolis algorithm\cite{METRO}.
Let us consider a probability distribution $P(x)$ from which we would like to sample data x.  
The MH algorithm consists of the following steps.

(1) First we set an initial value $x_0$ and $i=1$. 

(2) Then we generate a new value $x_i$ from a certain probability distribution $g(x_i|x_{i-1})$ 
which we call proposal density.    

(3) We accept the candidate $x_i$ with a probability of $P_{MH}(x_{i-1},x_i)$ 
where
\be
P_{MH}(x_{i-1},x_i) = \min\left[1,\frac{P(x_i)}{P(x_{i-1})}\frac{g(x_i|x_{i-1})}{g(x_{i-1}|x_i)}\right].
\label{eq:MH}
\ee
When $x_i$ is rejected we keep $x_{i-1}$, i.e. $x_i=x_{i-1}$.

(4) Go back to (2) with an increment of $i=i+1$.

For a symmetric proposal density $g(x_i|x_{i-1} )=g(x_{i-1}|x_i)$, eq.(\ref{eq:MH}) reduces to the Metropolis accept probability:
\be
P_{Metro}(x_{i-1},x_i)= \min\left[1,\frac{P(x_i)}{P(x_{i-1})}\right].
\ee

\section{Adaptive construction scheme}
Since the proposal density $g(x_i|x_{i-1})$ is dependent of the previous value $x_{i-1}$,
usually the sampled data are correlated.
One may use an independent proposal density $g(x_i)$ which does not depend on 
the previous value. Although in this case we can generate independent candidates $x_i$, 
it is important to choose the one close enough to the posterior density, 
in order to make the acceptance high enough.  

The posterior density of GARCH parameters often resembles to a Gaussian-like shape. 
Thus one may choose a density similar to a Gaussian distribution as the proposal density.  
Following \cite{WATANABE,ASAI},
in order to cover the tails of the posterior density
we use a (p-dimensional) multivariate Student's t-distribution given by 
\be
g(\theta)=\frac{\Gamma((\nu+p)/2)/\Gamma(\nu/2)}{\det \Sigma^{1/2} (\nu\pi)^{p/2}}
\left[1+\frac{(\theta-M)^t \Sigma^{-1}(\theta-M)}{\nu}\right]^{-(\nu+p)/2},
\label{eq:ST}
\ee
where $\theta$ and $M$ are column vectors,  
\be
\theta=\left[
\begin{array}{c}
\theta_1 \\
\theta_2 \\
\vdots \\
\theta_p
\end{array}
\right],
M=\left[
\begin{array}{c}
M_1 \\
M_2 \\
\vdots \\
M_p
\end{array}
\right],
\ee
and $M_i=E(\theta_i)$.
$\dis \Sigma$ is the covariance matrix defined as
\be
\frac{\nu\Sigma}{\nu-2}=E[(\theta-M)(\theta-M)^t].
\ee
For later use we also define a matrix $V$ as
\be
V=E[(\theta-M)(\theta-M)^t].
\ee
$\nu$ is a parameter to tune the shape of Student's t-distribution. 
When $\nu \rightarrow \infty$ the Student's t-distribution goes to a Gaussian distribution.
In this study we take $\nu=10$.

There are three parameters to be inferred for the GARCH(1,1) model. 
Therefore in this case $p=3$ and $\dis \theta=(\theta_1,\theta_2,\theta_3)=(\alpha,\beta,\omega)$, 
and $\Sigma$ is a $3\times3$ matrix.
The values of $\Sigma$ and $M$ are not known a priori.
We determine these unknown parameters $M$ and $\Sigma$ through MCMC simulations.
First we make a short run by the Metropolis algorithm and accumulate some data.
Then we estimate $M$ and $\Sigma$. Note that there is no need to estimate $M$ and $\Sigma$ accurately. 
Second we perform an MH simulation with the proposal density of eq.(\ref{eq:ST}) with the estimated $M$ and $\Sigma$.
After accumulating more data, we recalculate $M$ and $\Sigma$, and update $M$ and $\Sigma$ of eq.(\ref{eq:ST}).
By doing this, we adaptively change the shape of eq.(\ref{eq:ST}) to fit the posterior density more accurately.
We call eq.(\ref{eq:ST}) with the estimated $M$ and $\Sigma$ "adaptive proposal density".

The random number generation for the multivariate Student's t-distribution
can be done easily as follows.
First we decompose the symmetric covariance matrix $\Sigma$ by the Cholesky decomposition as
$\Sigma=LL^t$.
Then substituting this result to eq.(\ref{eq:ST}) we obtain
\be
g(X) \sim 
\left[1+\frac{X^t X}{\nu}\right]^{-(\nu+p)/2},
\ee
where $X=L^{-1}(\theta-M)$.
The random numbers $X$ are given by $\dis X=Y\sqrt{\frac{\nu}{w}}$,
where $Y$ follows $N(0,I)$ and $w$ is taken from the chi-square distribution $\nu$ degrees of freedom $\chi^2_{\nu}$.  
Finally we obtain the random number $\theta$ by $\theta=LX +M$.

\section{Empirical analysis}

We make an empirical analysis based on daily data of the exchange rates 
for US Dollar and Japanese Yen.
The sampling period of the exchange rates
is  4 January 1999  to 29 December 2006, which gives 2006 observations.
The exchange rates $p_i$ are transformed to $\dis r_i=100[\ln(p_i/p_{i-1})-\bar{s}]$
where $\dis \bar{s}$ stands for the average value of $\ln(p_i/p_{i-1})$.

Our implementation of the adaptive construction scheme is as follows.
First we make a short run by the Metropolis algorithm.
We discard the first 3000 data as burn-in process. 
Then we accumulate 1000 data to estimate $M$ and $\Sigma$. 
The estimated $M$ and $\Sigma$ are substituted to $g(\theta)$ of eq.(\ref{eq:ST}).
The shape parameter $\nu$ is set to 10.
We re-start a run by the MH algorithm with the proposal density $g(\theta)$. 
Every 1000 update we re-calculate $M$ and $\Sigma$ using  all accumulated data 
and update $g(\theta)$ for the next run.
We accumulate 100000 data for analysis.

\begin{table}[h]
  \centering
  \caption{Results of parameters.}
  \label{tab:1}
  {\footnotesize
    \begin{tabular}{clll}
      \hline
        & \multicolumn{1}{c}{$\alpha$} &
      \multicolumn{1}{c}{$\beta$} &
      \multicolumn{1}{c}{$\omega$} \\
      \hline
\hline
   Adaptive construction          & 0.03151 & 0.9403 & 0.01104  \\
   standard deviation\hspace{3mm} & 0.0078 & 0.017  & 0.0047 \\
   statistical error              & 0.00004 & 0.0001 & 0.00003 \\
   $2\tau_{int}$                  & $2.8 \pm 0.3$    & $3.8 \pm 0.4$ & $4.1\pm 0.5$     \\
\hline
   Metropolis          & 0.0318  & 0.9391 & 0.0114   \\
   standard deviation  & 0.0079  & 0.018  & 0.005 \\
   statistical error   & 0.0005  & 0.0014 & 0.0004 \\
   $2\tau_{int}$              & $400\pm 60$\hspace{2mm}   & $650\pm 100$\hspace{2mm}  & $620\pm 80$\hspace{2mm}   \\
      \hline
    \end{tabular}
  }
\end{table}

We also make a Metropolis simulation
and accumulate 100000 data for analysis.
The Metropolis algorithm in this study is implemented as follows.
We draw a new $\theta^\prime$ by 
adding a small random value $\delta \theta$ to the present value $\theta=(\theta_1,\theta_2,\theta_3)=(\alpha,\beta,\omega)$:
\be
\theta^\prime_j = \theta_j + \delta \theta_j,
\ee
where $\dis \delta \theta_j= d(r-0.5)$. 
$r$ is a uniform random number in $[0,1]$ and 
$d$ is a constant to tune the Metropolis acceptance.
We choose $d$ so that the acceptance becomes greater than $50\%$.

\begin{figure}
\vspace{5mm}
\centering
\includegraphics[height=4.1cm]{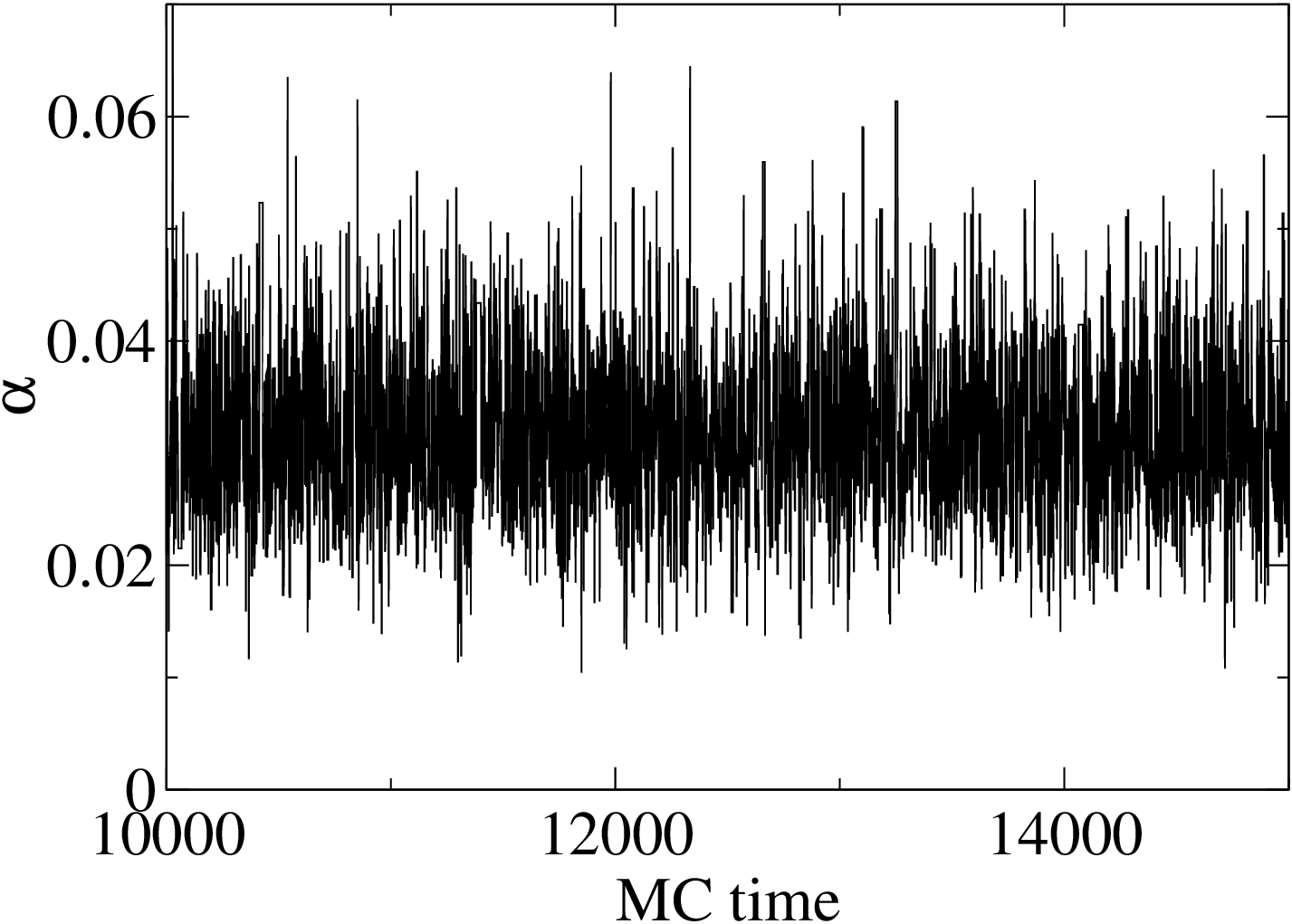}
\includegraphics[height=4.1cm]{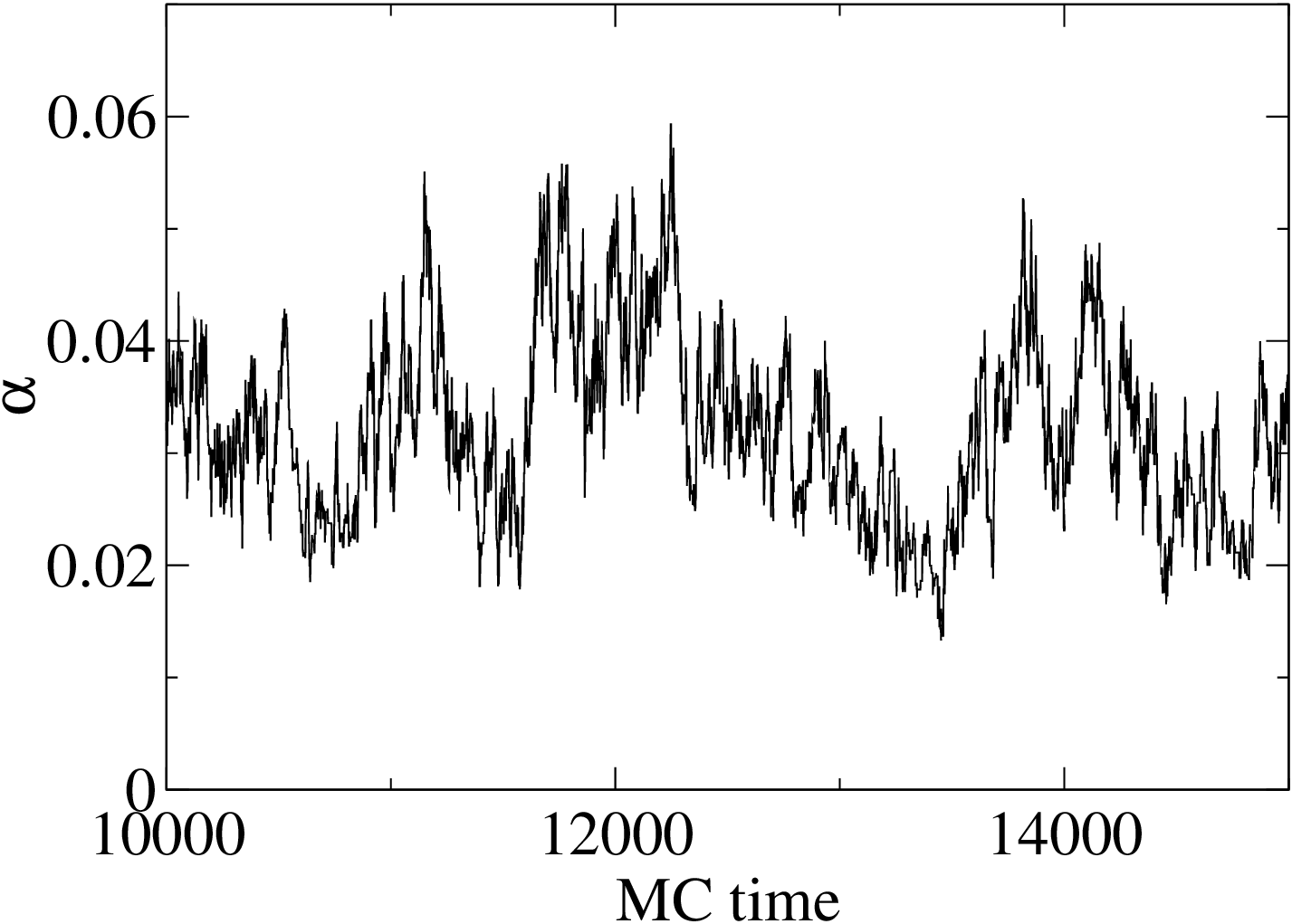}
\caption{
Monte Carlo time histories of $\alpha$ sampled by the adaptive construction scheme (left) 
and the Metropolis algorithm(right).
}
\vspace{1mm}
\label{fig:History}
\end{figure}

Fig.~1 compares  the Monte Carlo time history of $\alpha$ sampled by the adaptive construction scheme 
with that by the Metropolis algorithm.
It is clearly seen that the data $\alpha$ produced by the Metropolis algorithm are very correlated.
On the other hand the sampled data by the adaptive construction scheme seem to be well de-correlated. 
For other parameters $\beta$ and $\omega$ we also see the similar behavior.

In order to see correlations between sampled data, we measure the autocorrelation function (ACF)
defined as  
\be
ACF(t) = \frac{\frac1N\sum_{j=1}^N(x(j)- \bra x\ket )(x(j+t)-\bra x\ket)}{\sigma^2_x},
\ee
where $\bra x\ket$ and $\sigma^2_x$ are the average value and the variance of certain successive data $x$ respectively.

\begin{figure}
\vspace{5mm}
\centering
\includegraphics[height=4.0cm]{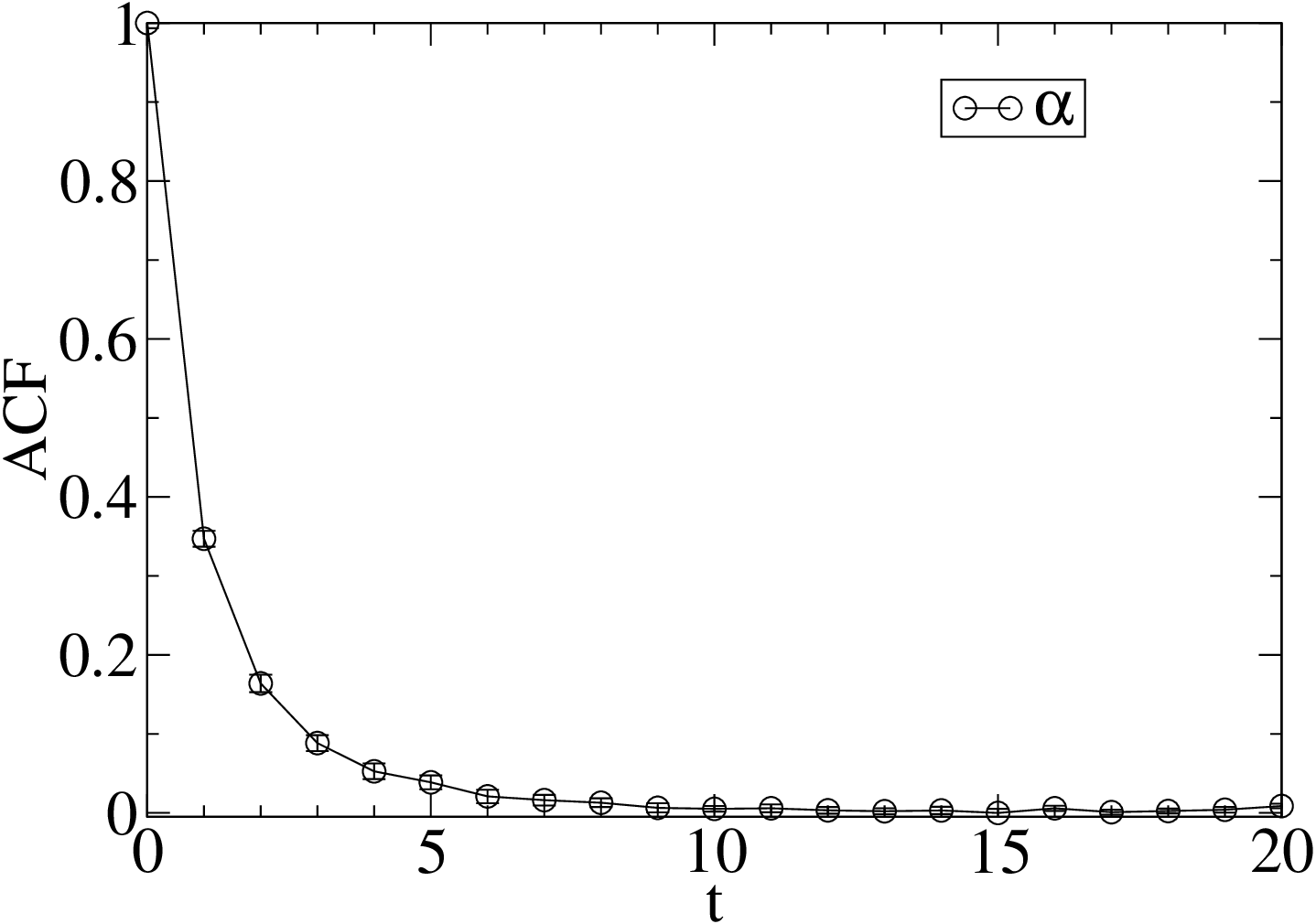}
\includegraphics[height=4.0cm]{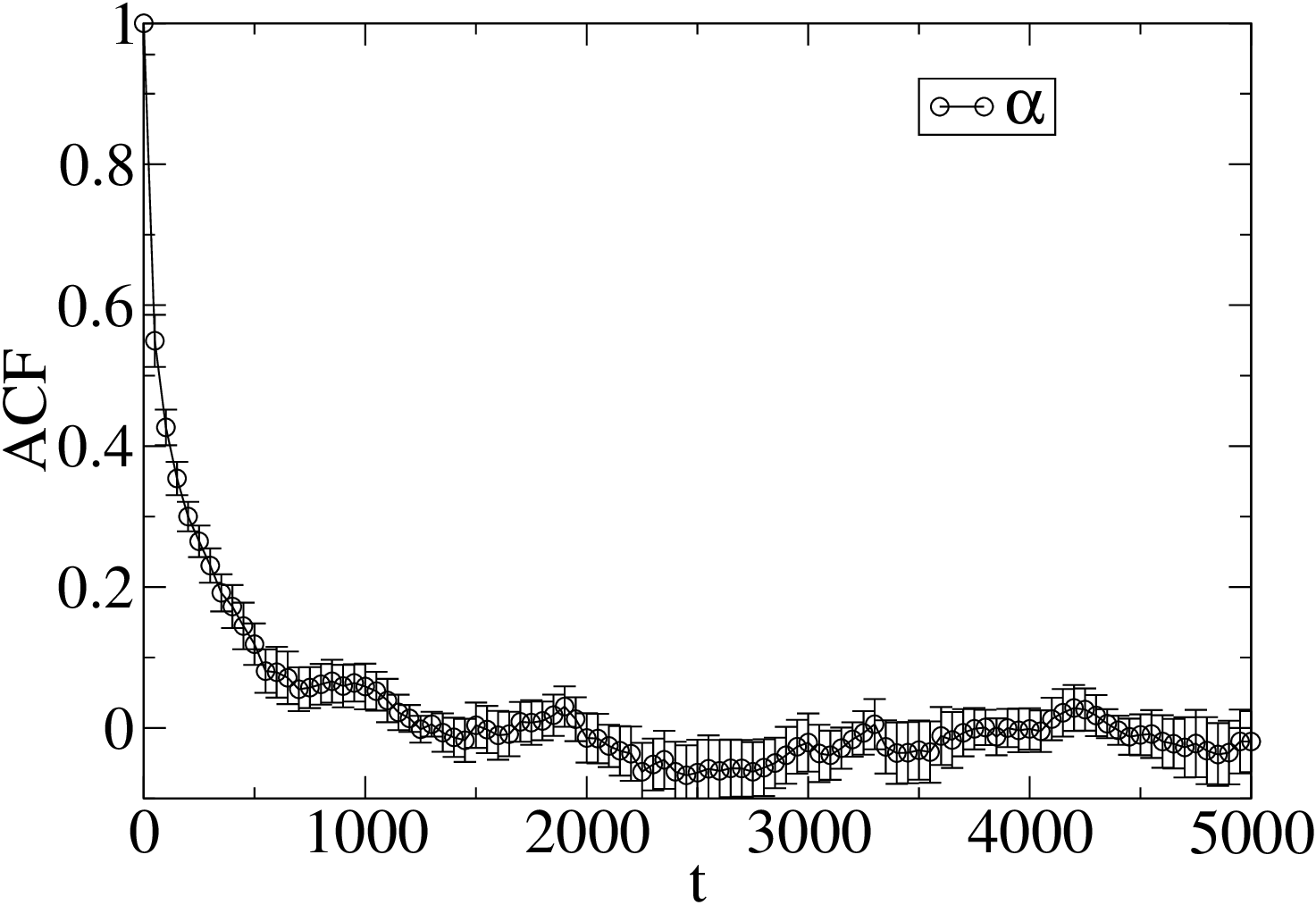}
\caption{
Autocorrelation functions of $\alpha$ for the adaptive construction scheme (left) and the Metropolis algorithm (right).
}
\vspace{1mm}
\label{fig:ACF}
\end{figure}

Fig.~2 shows the ACF for the adaptive construction scheme and the Metropolis algorithm.
The ACF of the the adaptive construction scheme decreases quickly as Monte Carlo time $t$ increases.
On the other hand the ACF of the Metropolis algorithm decreases very slowly which indicates that
the correlation between the sampled data is very large.

We estimate the autocorrelation time by the integrated autocorrelation time $\tau_{int}$.
To calculate $\tau_{int}$
we define $\tau_{int}(T)$ as
\be
\tau_{int}(T) = \frac12 +\sum_{i=1}^{T}ACF(i).
\ee
$\tau_{int}$ is given by  $\tau_{int} \def \tau_{int}(T=\infty)$.
In practice, however, it is impossible to sum up $ACF(t)$ to $T=\infty$.
Since typically $\tau_{int}(T)$ increases with $T$ and 
reaches a plateau 
we estimate $\tau_{int}$ at this plateau. 
Fig.~3 illustrates $\tau_{int}(T)$ of $\alpha$ sampled by the adaptive construction scheme.
$\tau_{int}(T)$ increases with $T$ and reaches a plateau around $T\ge 20 $.

\begin{figure}
\vspace{5mm}
\centering
\includegraphics[height=5.0cm]{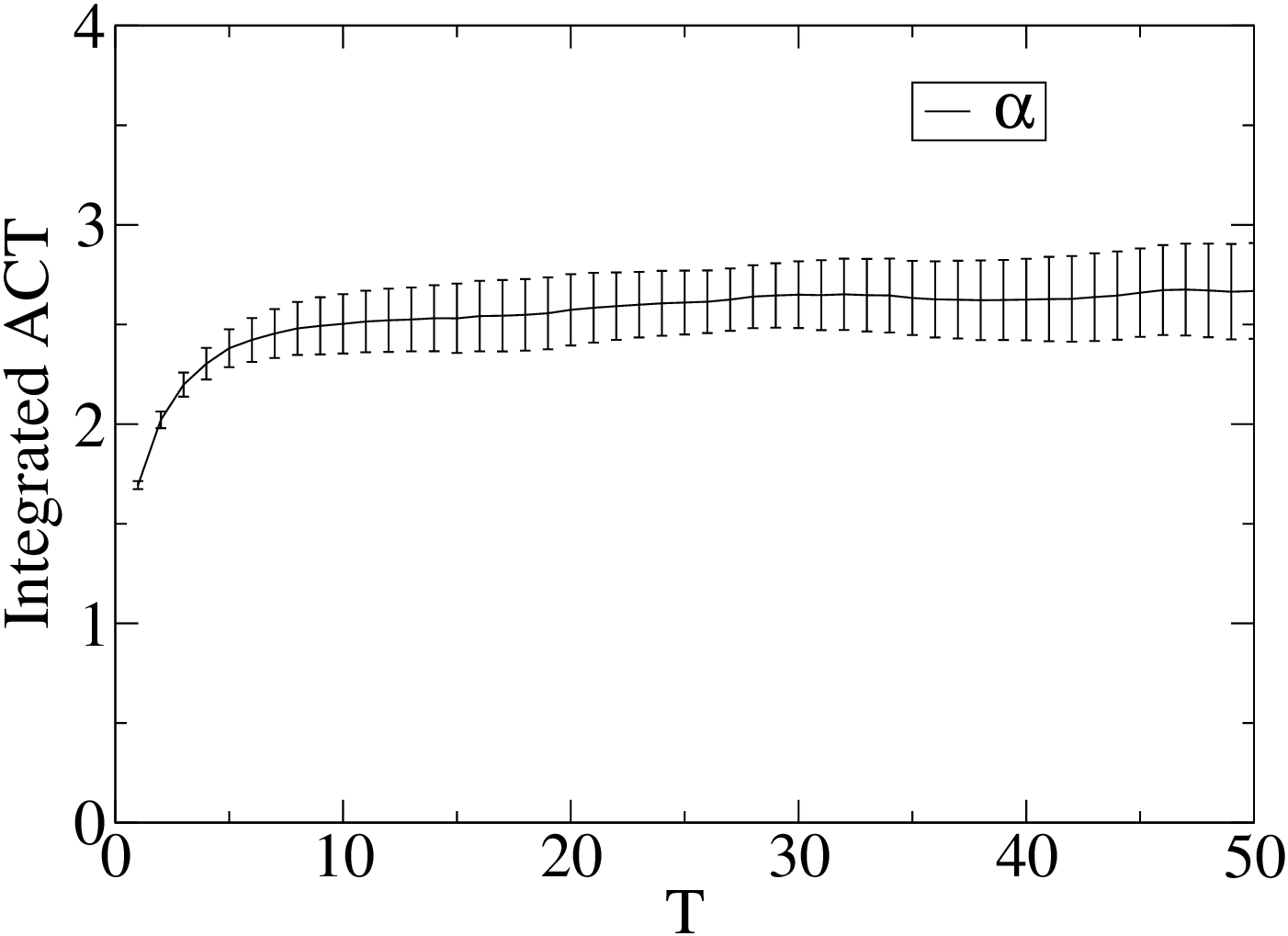}
\caption{
The integrated autocorrelation time $\tau_{int}(T)$ of $\alpha$ sampled by the adaptive construction scheme. 
}
\vspace{1mm}
\label{fig:ACT}
\end{figure}

Results of $\tau_{int}$ are summarized in Table 1. 
The values of $\tau_{int}$ from the Metropolis simulations 
are very large, typically several hundreds.
On the other hand $\tau_{int}$ from the adaptive construction scheme are very small, 
$2\tau_{int} \sim 2-3$\footnote{$2\tau_{int}$ is called an inefficiency factor.}.
This results in a factor of 10 reduction in terms of the statistical error. 
This reduction property is confirmed by the statistical errors of the sampled data (See Table 1). 
Thus it is concluded that the adaptive construction scheme is effectively working for 
reducing the correlations between the sampled data.

\begin{figure}
\vspace{5mm}
\centering
\includegraphics[height=4.0cm]{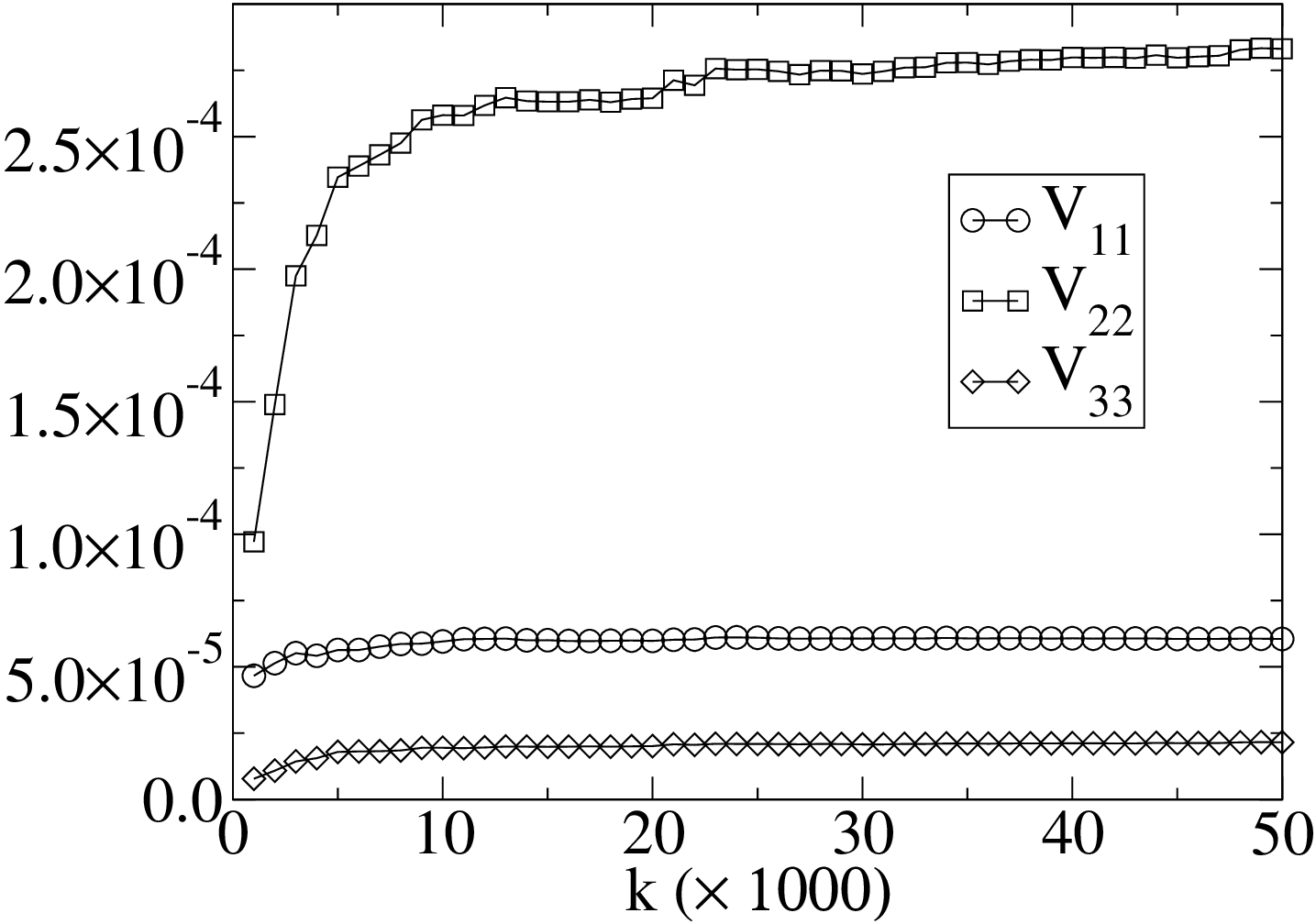}
\hspace{1mm}
\includegraphics[height=4.0cm]{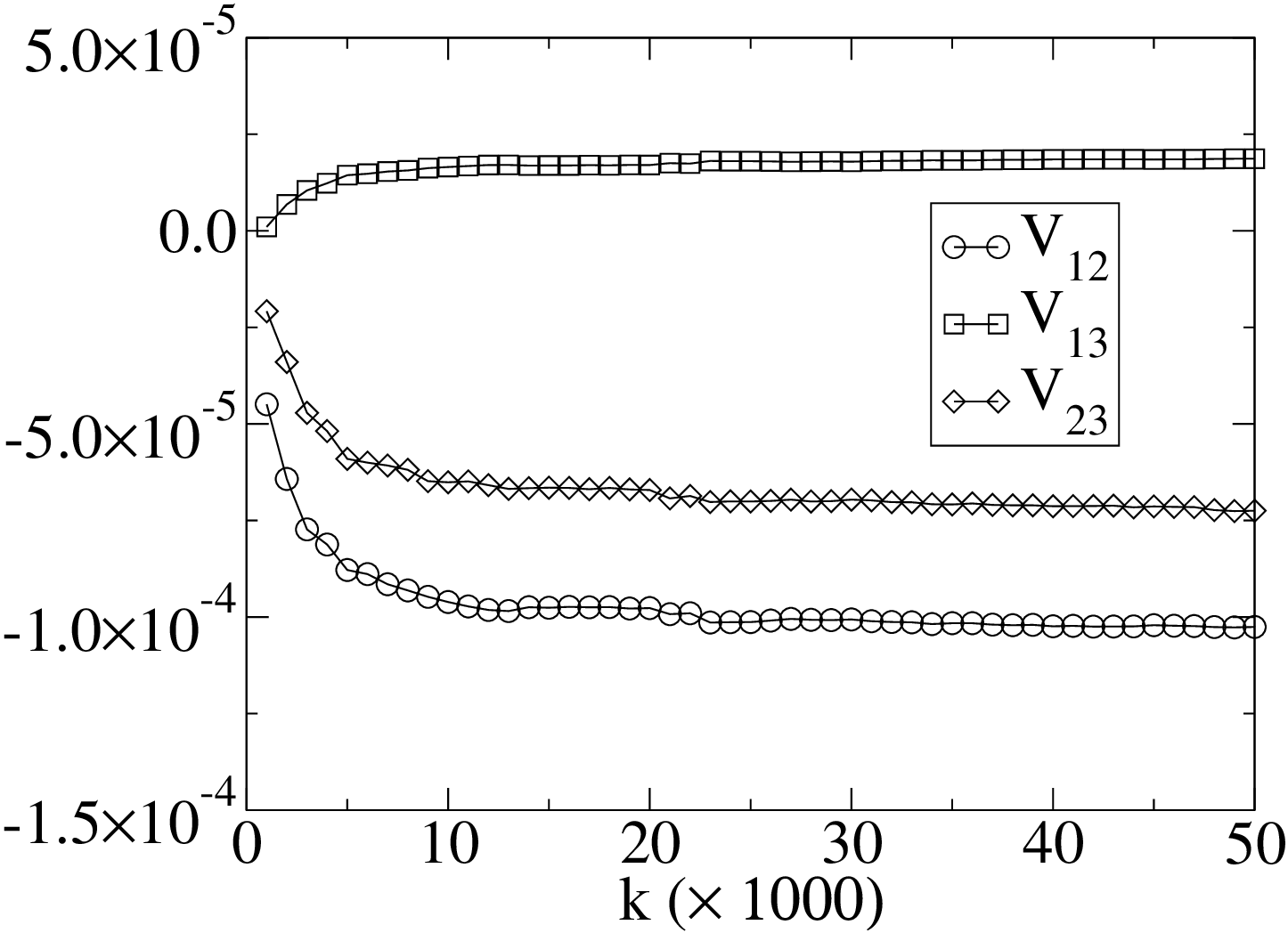}
\caption{
The matrix elements of the symmetric covariance matrix $V$. Diagonal elements (left) and off-diagonal elements (right).
}
\vspace{1mm}
\label{fig:SIG}
\end{figure}

Fig.~4 shows the convergence property of the matrix $V$.
The matrix elements $V_{ij}$ are defined by $\dis V=E[(\theta-M)(\theta-M)^t]$ 
with $\theta=(\theta_1,\theta_2,\theta_3)=(\alpha,\beta,\omega)$. For instance $V_{12}=V_{\alpha\beta}$.
All elements of $V$ converge quickly to certain values
as the simulations are proceeded.

\begin{figure}
\vspace{5mm}
\centering
\includegraphics[height=5.5cm]{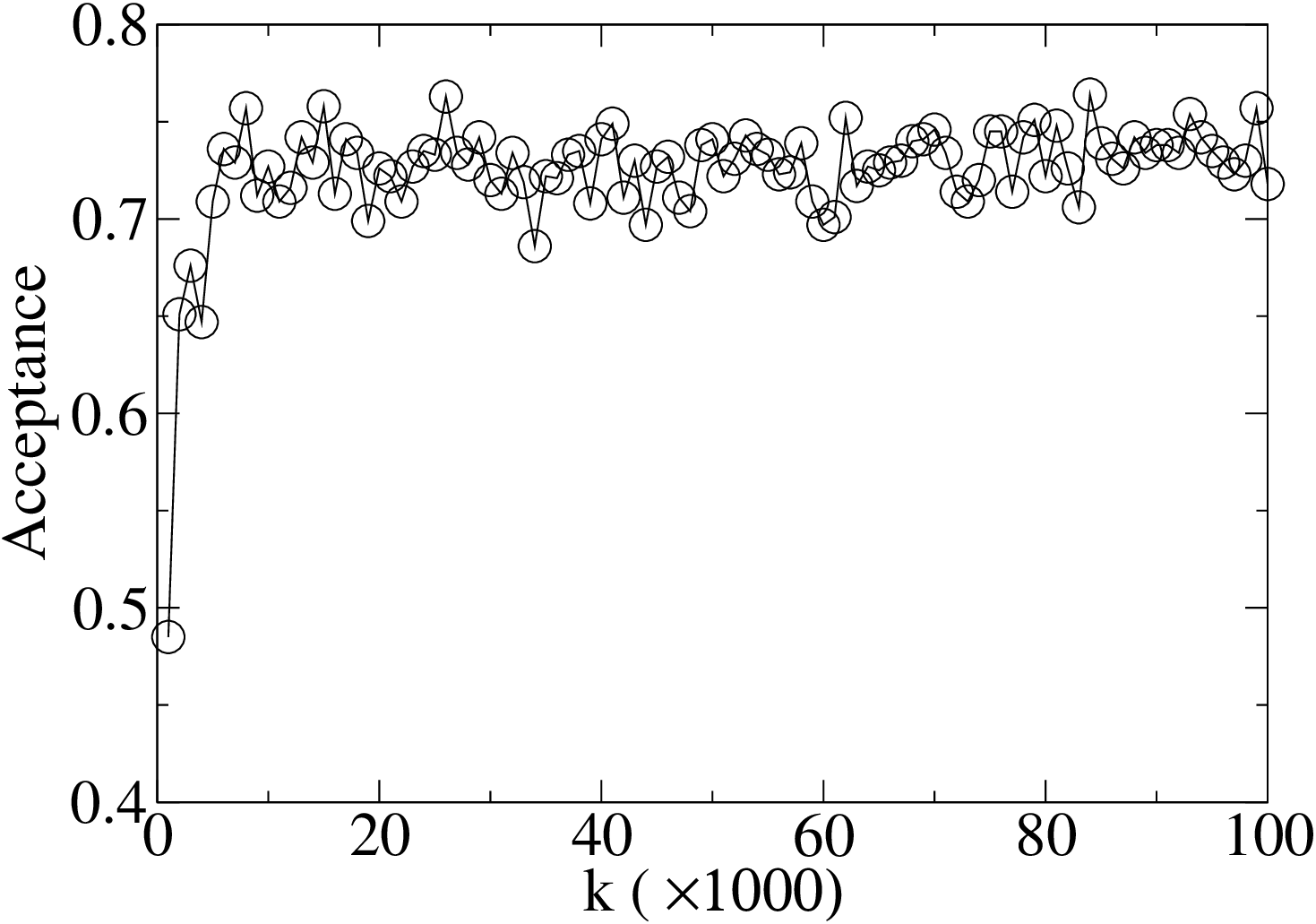}
\caption{
Acceptance at MH step with the adaptive proposal density.
}
\vspace{1mm}
\label{fig:ACC}
\end{figure}

Fig.~5  shows the acceptance at the MH algorithm with the adaptive proposal density of eq.(\ref{eq:ST}).
Each acceptance is calculated every 1000 updates and the calculation of the acceptance is
based on the latest 1000 data.
At the first stage of the simulation
the acceptance is low. This is because at this stage $M$ and $\Sigma$ are not calculated accurately yet.
However the acceptances increase quickly as the simulations are proceeded
and reaches a plateau where the acceptance is more than 70\%.

\section{Summary}
We proposed the adaptive construction scheme to construct a proposal density for the MH algorithm of the GARCH(1,1) model. 
The construction of the proposal density is performed using the data generated by MCMC methods.
During the MCMC simulations the proposal density is updated adaptively.
In this study we applied the adaptive construction scheme for the Bayesian inference of
the GARCH(1,1) model by using US Dollar/Japanese Yen exchange rate.
The numerical results show 
that the adaptive construction scheme significantly reduces the correlations between the sampled data.
The autocorrelation time of the adaptive construction method is calculated to be $2\tau_{int}\sim 2-3$, 
which is comparable to that of the AR/MH method\cite{ASAI}.
It is concluded that the adaptive construction scheme is an efficient method 
for the Bayesian inference of the GARCH(1,1) model.
The adaptive construction scheme is not limited to the GARCH(1,1) model and can be applied for other GARCH-type models.

\section*{Acknowledgments}
The numerical calculations were carried out on Altix at the Institute of Statistical Mathematics
and on SX8 at the Yukawa Institute for Theoretical Physics 
in Kyoto University.

\end{document}